\documentclass[english]{article}
\usepackage[T1]{fontenc}
\usepackage[latin9]{inputenc}
\usepackage{geometry}
\usepackage{float}
\usepackage{amsmath}
\usepackage{graphicx}
\usepackage{setspace}
\makeatletter
\newcommand{\lyxaddress}[1]{
	\par {\raggedright #1
	\vspace{1.4em}
	\noindent\par}
}

\makeatother

\usepackage{babel}
\begin{document}
\title{Efficient and flexible methods for time since infection models}
\author{Joseph D. Peterson and Ronojoy Adhikari}
\maketitle

\lyxaddress{\begin{center}
University of Cambridge, DAMTP, Center for Mathematical Sciences, Wilberforce Rd., Cambridge CB3 0WA, United Kingdom
\par\end{center}}
\begin{abstract}
Epidemic models are useful tools in the fight against infectious diseases, as they allow policy makers
to test and compare various strategies to limit disease transmission while mitigating collateral damage
on the economy. Epidemic models that are more faithful to the microscopic details of disease transmission
can offer more reliable projections, which in turn can lead to more reliable control strategies. For
example, many epidemic models describe disease progression via a series of artificial 'stages' or 'compartments'
(e.g. exposed, activated, infectious, etc.) but an epidemic model that explicitly tracks time since infection
(TSI) can provide a more precise description. At present, epidemic models with 'compartments' are more
common than TSI models , largely due to the higher computational cost and complexity typically associated
with TSI models. Here, however, we show that with the right discretization scheme a TSI model is not
much more difficult to solve than a comparment model with three or four 'stages' for the infected class.
We also provide a new perspective for adding 'stages' to a TSI model in a way that decouples the disease
transmission dynamics from the residence time distributions at each stage. These results are also generalized
for age-structured TSI models in an appendix. Finally, as proof-of-principle for the efficiency of the
proposed numerical methods, we provide calculations for optimal epidemic control by non-pharmaceutical
intervention. Many of the tools described in this report are available through the software package 'pyross'
\end{abstract}

\section{Background and Introduction}

Epidemic modelling is a valuable tool for both (1) predicting how a disease will spread through a population
and (2) testing and optimizing control strategies to mitigate the damage. Better control strategies are
possible with better epidemic modelling, and ultimately an epidemic model is only as reliable as the
assumptions it makes. Good epidemic models must be able to provide fine-grained descriptions for how
a disease is spreading. It is not enough to simply predict a number of infected persons -- one must
also be able to predict the spread of disease from city-to-city and age group to age group, for example
\cite{key-4,key-5,key-6,key-7}. Given the full burden of complexity an epidemic model is expected to
bear, one cannot usually afford much more than an elementary model of disease transmission at each level.

To that end, some of the most popular classes of epidemic models are generalizations of the standard
SIR framework, in which a population is partitioned into discrete compartments of susceptibles (S), infecteds
(I), and recovered (R) \cite{key-8}. In the simplest instance of this framework, a susceptible individual
becomes infected by encountering an infected person in some way or another. An infected person is immediately
infectious, and the subsequent transition to recovered has no memory of the time spent being infected
(i.e. it is a random process with an exponential distribution of residence times). In the limit of a
large population, the mean result over a large number of random transitions is predictable, and so deterministic
versions of the model (like those considered in this report) are often preferred. Compartment-based models
like SIR model do capture important features of disease transmission, but the reliance on an artificial
construction of discrete stages with Markov transitions is at odds with the complex medical realities
of disease progression. This can cause problems, for example, when predicting the reproduction number
of a disease based on its initial growth dynamics \cite{key-10}.

A more realistic model of disease transmission describes the infectiousness of an infected individual as
a function of the infection age, or the 'time since infection' (TSI) \cite{key-8,key-5,key-9,key-1,key-2}
Does a person become infectious right away, or is there a delay of several days before they can transmit?
When is a person most infectious? What is the probability distribution for recovery times? These details
(which define the actual transmission dynamics) are naturally incorporated into a TSI model, but they
can be difficult to match precisely in a compartment model \cite{key-10,key-11}. On the surface, it
might seem that any epidemic model based on time since infection will be substantially more complex than
the basic SIR model and therefore of limited value to fine-grained epidemic modelling applications. In
this report, however, we will show that the SIR model can be extended to consider time-since-infection
without making the model much more difficult to solve.

The advantages of a TSI model for disease dynamics do not eliminate the need to resolve discrete compartments.
For example, in modeling the transmission of SARS-CoV-2, in addition to knowing the total number of infected
at any point in time, one must also make predictions for the total number of hospitalizations if there
is a risk that the healthcare system might become overwhelmed \cite{key-3,key-14}. In this report, we
present a very simple and flexible strategy for decoupling the residence time distributions at each stage
from the overall dynamics for transmission. Instead of explicitly modelling the occupancy and infectiousness
in each compartment, we simply keep track of the total infected population at every TSI and then apply
a filter to determine the total occupancy of each compartment. This makes the whole modelling framework
more flexible and it should also be more straightforward to parameterize from case data.

By way of overview, this report is arranged as follows: First, in section \ref{sec:Governing-Equations},
we present the governing equations for a TSI model, including the use of a filter over the infected population.
We non-dimensionalize the governing equations (section \ref{subsec:Nondimensionalization}) and then
offer three different numerical methods by which the TSI model can be integrated in time. In particular,
section \ref{subsec:Galerkin} introduces a novel Galerkin approximation with a spectral discretization
of the infected population. The generalization of section \ref{sec:Governing-Equations} to age-structured
epidemic models is provided in the appendix. In section \ref{sec:Sample-Results:-Simple}, we provide
sample results for epidemic curves obtained from the three discretization strategies, and in section
\ref{sec:Galerkin_problems}, we discuss potential limitations of the Galerkin discretization. Finally,
in section \ref{sec:Sample-Results:-Optimal}, we go beyond simple epidemic curves to consider optimal
control by non-pharmaceutical interventions. To our knowledge, this is the first time that such calculations
have been performed for a TSI modelling framework.

The methods described in the present report are available for use in the pyRoss epidemic modelling package
(https://github.com/rajeshrinet/pyross).

\section{\label{sec:Governing-Equations}Governing Equations}

\subsection{The deterministic SIR model}

For pedagogical purposes, we first review the deterministic SIR model. The SIR model is simpler than
the basic TSI model, but it shares a similar underlying structure. The number of susceptibles, $S$,
infecteds, $I$, and recovereds, $R$ evolve in time by:

\begin{equation}
\frac{dS}{dt}=-\frac{\beta}{N}IS
\end{equation}

\begin{equation}
\frac{dI}{dt}=\frac{\beta}{N}IS-\gamma I
\end{equation}

\begin{equation}
\frac{dR}{dt}=\gamma I
\end{equation}
where $\beta$ and $\gamma$ are rate constants for infection and recovery, respectively (units of 1/time)
and $N=S+I+R$ is the total population.

\subsection{Basic deterministic TSI equations}

\begin{figure}[H]
\centering{}\includegraphics[scale=0.3]{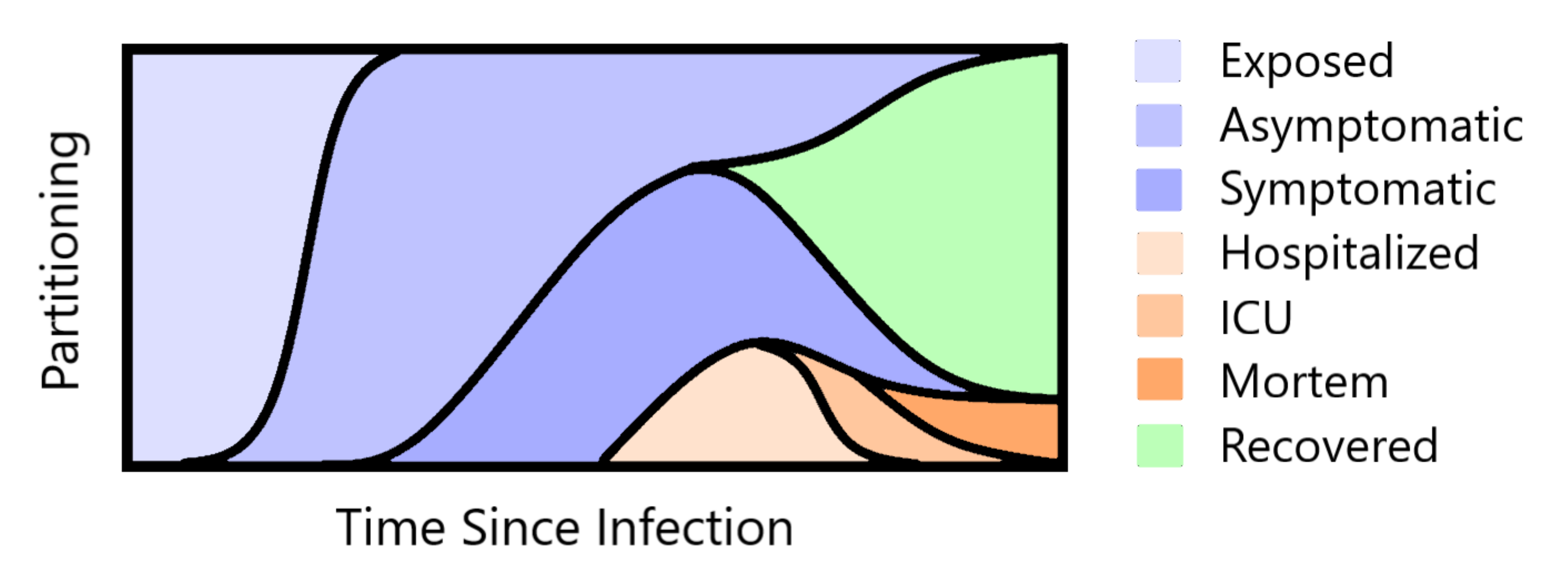}\caption{\label{fig:partitioning}A graphical representation of how an infected population might be partitioned
into different categories.}
\end{figure}

Models that include time since infection are not new, and in fact the equations presented here (apart
from superficial differences pertaining to the filtering convention, c.f. Figure \ref{fig:partitioning})
are covered by existing models for epidemics, many of which are more general \cite{key-5,key-9,key-12}.
Historically, however, models with TSI have been under-utilized due to the higher computational cost
and complexity associated therewith. It should also be noted that compartment models like SIR can also
be represented as TSI models \cite{key-15} and TSI models can also be discretized into compartment models
(c.f. section \ref{subsec:SIkR}). The distinction between the two is one of frameworks, not physics.
When it is important to accurately describe disease dynamics in terms of time since infection, it is
our view that the TSI modelling framework provides a more efficient and flexible set of tools to work
with.

In a model with time since infection, we will speak of the infected class in terms of the number density
of persons (per unit time) whose infections began at a time s prior to the current time $t$, $I(t,s)$.
Thus, the total number of persons infected in the narrow interval between $s$ and $s+\delta s$ prior
to time $t$ is given by $I(t,s)\delta s$. As time passes, the time since infection for all infected
persons also continues to pass in the same way. Given $I(t,s)$, one can calculate the rate at which
susceptible persons become infected:

\begin{equation}
\frac{dS}{dt}=-\int_{0}^{T}ds\frac{\beta(s)}{N}I(t,s)S\label{eq:SIcR_begin}
\end{equation}
Note that the mean rate constant for infection $\beta(s)$ now varies with the time since infection.
Here we have assumed that infections older than time $T$ are no longer transmitting, $\beta(s>T)=0$,
which allows the integral to be truncated at time $T$. We have also ignored the possibility of re-infection
for the time being. The number density equation for the infected class evolves in time due to the passage
of time,

\begin{equation}
\frac{\partial}{\partial t}I(t,s)+\frac{\partial}{\partial s}I(t,s)=0
\end{equation}
and the number density of new infections must match the rate at which susceptible individuals are becoming
infected:

\begin{equation}
I(t,0)=\int_{0}^{T}ds\frac{\beta(s)}{N}I(t,s)S
\end{equation}

As with the standard SIR model, the equations for susceptibles and infecteds are closed. However, for
practical applications of an epidemic model one may also need to describe the various sub-classes of
the infected population (e.g. infectious, recovered, hospitalized, deceased, quarantined, asymptomatic,
etc), with each sub-class $\alpha$ representing a fraction $\Phi^{\alpha}(s)$ of the total with number
density $I^{\alpha}(t,s)=\Phi^{\alpha}(s)I(t,s)$ . The total population in each sub-class at any time
is then given by:

\begin{equation}
I^{\alpha}(t)=\int_{0}^{\infty}ds\Phi^{\alpha}(s)I(t,s)\label{eq:SIcR_end}
\end{equation}
This partitioning, or 'filtering', is also represented graphically in Figure \ref{fig:partitioning}.

The partitioning strategy is somewhat different than what one often sees in a TSI model containing sub-classes
of the infected population, and it is only viable because we are interested in the deterministic (as
opposed to stochastic) version of the model. To our understanding, the more common treatment (for both
deterministic and stochastic models) is to explicitly move the infected population between different
compartments while keeping track of the 'time since arrival' in each compartment \cite{key-3,key-15}.
We prefer the partitioning scheme presented here for two reasons: First, it eliminates the need for and
the influence of artificially constructed stages (e.g. 'exposed', 'asymptomatic', etc.). Second, it decouples
assignments to compartments from the disease dynamics defined by $\beta(s).$ Third, and perhaps most
importantly, it is a severe approximation to say that the distribution of exit times from a present stage
(e.g. hospitalization) depends only on the time since arrival and not on the time since infection. In
our approach, we note that since time since arrival depends on time since infection, the overall entry/exit
rates for hospitaization can be exactly described as a function of time since infection with no approximations
necessary.

For many compartment models (and indeed many TSI models as well) it is not unusual to suppose that infectiousness
of an individual depends on the subclass or subclasses to which the individual has been assigned \cite{key-14}.
In that case, if each sub-class has a rate constant for infection $\beta_{\alpha}(s)$ equation \ref{eq:SIcR_begin}
would look like:

\begin{equation}
\frac{dS}{dt}=-\sum_{\alpha}\int_{0}^{T}ds\frac{\beta_{\alpha}(s)}{N}I^{\alpha}(t,s)S\label{eq:SIcR_w_subclasses}
\end{equation}
In our construction of a TSI model, however, explicitly evaluating the summation is unnecessary -- we
have already stated that $\beta(s)$ is the mean infectiousness of the infected class at time since infection
$s$. In other words, equation \ref{eq:SIcR_w_subclasses} is equivalent to equation \ref{eq:SIcR_begin}
when the mean value $\beta(s)$ is correctly defined as:

\begin{equation}
\beta(s)=\sum_{\alpha}\Phi^{\alpha}(s)\beta_{\alpha}(s)
\end{equation}
Thus, we can independently control the overall disease dynamics given by $\beta(s)$ and the overall
partitioning given by $\Phi^{\alpha}$ without having to settle the details for each $\beta_{\alpha}(s)$.

For the analysis in this report, we will only present calculations for two subclasses of infecteds --
those who are recovered, with population $R$, and those that are not. If all infected persons are delcared
recovered only once their infection has aged beyond the longest possible transmission time, then $\Phi_{R}(s)$
is just a step function at $s=T$. Taking a time derivative of equation \ref{eq:SIcR_end}, this can
also be written as:

\begin{equation}
\frac{dR}{dt}=I(t,T)
\end{equation}
Or, more generally, given $\phi^{\alpha}(s)=d\Phi^{\alpha}/ds$ the population in any subclass of infecteds
evolves by:

\begin{equation}
\frac{dI^{\alpha}}{dt}=\int_{0}^{T}ds\phi^{\alpha}(s)I(t,s)
\end{equation}

In general, one can choose any number of subclasses with arbitrarily complex shapes for each $\Phi_{\alpha}(s)$
with minimal added cost for obtaining solutions numerically (provided one chooses an appropriate numerical
method). We refer to equations \ref{eq:SIcR_begin} - \ref{eq:SIcR_end} as the TSI model. Equations
for an age-structured generalization of this TSI model are provided in the appendix.

\subsection{Additional Remarks on Partitioning the Infecteds}

In equations \ref{eq:SIcR_begin} - \ref{eq:SIcR_end}, it was assumed that the functions $\Phi^{\alpha}$
partitioning the infecteds depend on time since infection but not on time itself. This approximation
works well for modelling subclasses whose populations reflect the nature of the disease itself (like
hospitalizations and fatalities), but it is less suited to subclasses whose populations reflect a behavioral
or societal response to the disease. Consider a system in which quarantines are assigned by a track-and-trace
progam -- the probability of being under quarantine at time since infection $s$ depends on the history
of testing to that point and not necessarily on the current state of testing. Therefore, in the most
general case one can write a dynamical equation:

\begin{equation}
\frac{\partial\Phi^{\alpha}}{\partial t}+\frac{\partial\Phi^{\alpha}}{\partial s}=\phi_{in}^{\alpha}(t,s)-\phi_{out}^{\alpha}(t,s)\label{eq:Phi_all}
\end{equation}
where $\phi_{in}^{\alpha}(t,s)$ and $\phi_{out}^{\alpha}(t,s)$ are the probability density for moving
in/out of subclass $\alpha$ at time $t$ and time since infection s. Note, however, that if the criterion
for moving people in and out of subclass $\alpha$ are changing on timescales that are slow compared
to the typical residence time for subclass $\alpha$, then to leading order one can apply a quasi-static
approximation for which equation \ref{eq:Phi_all} reverts to:

\begin{equation}
\Phi^{\alpha}(t,s)=\int_{0}^{s}ds'\phi^{\alpha}(t,s')
\end{equation}

\begin{equation}
\phi^{\alpha}(t,s)=\phi_{in}^{\alpha}(t,s)-\phi_{out}^{\alpha}(t,s)
\end{equation}
and hence equation \ref{eq:SIcR_end} becomes:

\begin{equation}
\frac{dI^{\alpha}}{dt}=\int_{0}^{T}ds\phi^{\alpha}(t,s)I(t,s)
\end{equation}

Finally, while the underlying probability density functions $\phi^{\alpha}(t,s)$ can be obtained from
track-and-trace data, we have found that the available data is not sufficiently detailed to precisely
specify all necessary inputs to a time since infection model. To that end, we suggest that the probability
density functions can be approximated as beta distributions:

\begin{equation}
\phi_{in}^{\alpha}(s)=p_{in}^{\alpha}\frac{1}{T}\frac{\Gamma(a+b)}{\Gamma(a)\Gamma(b)}\Big(\frac{s}{T}\Big)^{1-a}\Big(1-\frac{s}{T}\Big)^{1-b}
\end{equation}
where $p_{in}^{\alpha}$ is the probability that an infected person is ever assigned to subclass $\alpha$
and the 'shape' of the distribution is defined by parameters $a$ and $b$. The same parameterization
can be done for each $\phi_{out}^{\alpha}$ as well. Finally, the shape parameters $\alpha$ and $\beta$
can be explicitly related to the mean $\bar{x}$ and variance $v$ of the distribution:

\begin{equation}
a=\frac{\bar{x}^{2}(1-\bar{x})}{v}-\bar{x}
\end{equation}

\begin{equation}
b=\Big(\frac{\bar{x}(1-\bar{x})}{v}-1\Big)(1-\bar{x})
\end{equation}

This same parameterization strategy also applies to the rate constant for infection, $\beta(s)$, which
is itself a probability density function for transmitting the disease at time since infection $s$.

\subsection{\label{subsec:Nondimensionalization}Nondimensionalization}

Before proceeding further, it is helpful to recast the governing equations into dimensionless form. We
rescale $S$ and $R$ by the total population $N$, such that $\tilde{S}=S/N$ and $\tilde{R}=R/N$.
Using a characteristic time $T_{C}=T/2$, we rescale time as $\tilde{t}=t/T_{C}$ and number density
of infecteds as $\tilde{I}=IT_{C}/N$. The time since infection is rescaled to the variable $\tilde{s}=s/T_{C}-1$
so that the range $s\in[0,T]$ is mapped to $\tilde{s}\in[-1,1]$. Given the reproduction number $R_{0}=\int_{0}^{T}\beta(s)ds$
we rescale $\tilde{\beta}=\beta T_{C}/R_{\ensuremath{0}}$ so that $\int_{-1}^{1}d\tilde{s}\tilde{\beta}(\tilde{s})=1$,
and we rescale all $\tilde{\phi}^{\alpha}=\phi^{\alpha}T_{C}$ in the same way so that $\int_{-1}^{1}d\tilde{s}\tilde{\phi}^{\alpha}(\tilde{s})=p^{\alpha}$,
where $p^{\alpha}$ is the probability that $\alpha$ is the final state (e.g. recovered, deceased, etc)
for an infected person.

For compactness of notation, we will suppress tildes in our dimensionless variables -- from this point
forward, all variables are dimensionless unless otherwise stated. The dimensionless governing equations
of the TSI model with time since infection are given by:

\begin{equation}
\frac{dS}{dt}=-R_{0}S\int_{-1}^{1}ds\beta(s)I(t,s)\label{eq:ndim_begin}
\end{equation}

\begin{equation}
\frac{\partial}{\partial t}I(t,s)+\frac{\partial}{\partial s}I(t,s)=0\label{eq:ndim_I}
\end{equation}

\begin{equation}
I(t,-1)=R_{0}S\int_{-1}^{1}ds\beta(s)I(t,s)\label{eq:ndim_bc}
\end{equation}

\begin{equation}
\frac{dR}{dt}=I(t,s=1)
\end{equation}

\begin{equation}
\frac{dI^{\alpha}}{dt}=\int_{-1}^{1}ds\phi^{\alpha}(s)I(t,s)
\end{equation}
These equations can be solved by any number of means. In the section that follows, we will present three
discretization stategies for numerical solution. First, we show that existing multi-stage models can
be derived using a particularly inefficient method-of-lines discretization. Second, we show that with
a few small changes the accuracy of the standard multi-stage model can be improved dramatically at a
small cost -- time-stepping is inflexible. Finally, for both spectral accuracy in $s$ and flexible
time-stepping in $t$, we present a novel discretization based on a Galerkin approximation.

\subsection{\label{subsec:SIkR}Method of Lines: the SIkR model}

The PDE can be converted to an approximate ODE representation via the method of lines, perhaps the most
common technique used to solve TSI models \cite{key-15}. Here, we discretize $s$ in uniform intervals
of width $h$ and represent advection/integration in $s$ via upwinding and right Riemann sums, respectively.
These methods are robust but generally poor performing, with accuracy $O(h)$ - about as bad as any stable
numerical scheme can be. We will show that this choice of discretization maps the TSI model to the standard
multi-stage generalization of the SIR model \cite{key-13}. Details about the truncation errors introduced
by this particular choice of discretization will be discussed shortly.

We evaluate the solution at discrete points $s_{1},s_{2},\ldots,s_{N}$ where $s_{n}=-1+(n-1)h$ and
$h=2/(N-1)$. In this section, we denote $I_{n}$ as the number density of infected persons at each $s_{n}$
and $\beta_{n}\sim h\beta(s_{n})$. For best results, the value of $\beta_{n}$ should be re-scaled so
that the right Riemann sum approximation still evaluates to unity: $\sum_{n=2}^{N}\beta_{n}=1$. Applying
this discretization scheme to the PDE equations, we obtain evolution equations for $S,I_{n},$ and $R$:

\begin{equation}
\frac{dS}{dt}=-R_{0}S\sum_{n=2}^{N}\beta_{n}I_{n}\label{eq:S_SIkr}
\end{equation}

\begin{equation}
I_{1}=R_{0}S\sum_{n=2}^{N}\beta_{n}I_{n}
\end{equation}

\begin{equation}
\frac{dI_{n}}{dt}=\frac{1}{h}(I_{n-1}-I_{n})
\end{equation}

\begin{equation}
\frac{dR}{dt}=I_{N}\label{eq:R_SIkR}
\end{equation}

Recall, however, that equation \ref{eq:R_SIkR} applies to the special case where a person recovers only
after aging beyond the maximum infectious period, $\phi_{R}(s)=\delta(s-1)$. For more realistic versions
of $\phi_{R}(s)$, equation \ref{eq:R_SIkR} becomes:

\begin{equation}
\frac{dR}{dt}=\sum_{n=2}^{N}\phi_{R,n}I_{n}\label{eq:R_SIkR_general}
\end{equation}

\begin{equation}
\phi_{R,n}\sim h\phi_{R}(s_{n})
\end{equation}
For best results, the value of $\phi_{R,n}$ should be re-scaled so that the right Riemann integral still
evaluates to one, $\sum_{n=2}^{N}\phi_{R,n}=1$.

Equations \ref{eq:S_SIkr} - \ref{eq:R_SIkR} map exactly to the SIR model when $N=2$ or to a multi-stage
SIkR model with $k=N-1$ stages more generally. In practice, however, the convergence is so slow that
truncation errors will have a large influence on the predictions. Therefore, we discuss the truncation
errors introduced by this scheme and how they systematically influence the resulting predictions.

First, we consider truncation error when integrating over s using a left Riemann sum. Assuming $\beta_{n}=\beta_{0}$
for all n, the right Riemann sum will under/over estimate the infectivity of the infected class during
the rising/falling of the epidemic by $O(h)$. Next, upwinding to approximate derivatives in $s$ introduces
numerical diffusion, meaning that the numerical solution behaves like a PDE model to which an additional
diffusion term has been added:

\begin{equation}
\frac{\partial}{\partial t}I(t,s)+\frac{\partial}{\partial s}I(t,s)=D\frac{\partial^{2}}{\partial s^{2}}I(t,s)
\end{equation}

Where the numerical diffusion coefficient $D$ scales as $D\sim O(h)$. Thus, the upwinding scheme tends
to smear out the distribution of infection times. This error is less important if $\beta(s)$ is constant,
but in general the resulting numerical diffusion will tend to over-estimate the number of late/early
stage infected during the rising/falling of the epidemic by $O(h)$. Numerical diffusion can be eliminated
if the discretized ODE problem is integrated via forward Euler with a time-step of exactly $h$ -- this
moves the truncation error from the $s-$domain into the $t-$domain, but it keeps the overall error
at $O(h)$ and allows for faster calculation.

\subsection{\label{subsec:PC}Method of Lines: Predictor/Corrector}

In the preceding section, we showed that the SIkR model is an approximation of our TSI model with first
order accuracy, $O(h)$. The accuracy can be improved to second order $O(h^{2})$ with a few small modifications.
We use a trapezoid rule (as opposed to a right Riemann sum) for quadrature in $s$ and a predictor-corrector
midpoint rule for evolving the susceptible and recovered populations in time. The infected population
(with the exception of newly infecteds) are upwinded with a Courant number of one to eliminate numerical
diffusion. Similar methods have been previously reported for similar epidemic models \cite{key-2}. The
principle disadvantage of this method is that time-stepping is inflexible.

We evaluate the solution at discrete points $s_{1},s_{2},s_{3}\ldots,s_{N}$ where $s_{n}=-1+(n-1)h$
and $h=2/(N-1)$ and also at discrete times $t^{(k)}=kh$ for integer $k$. We denote $I_{n}$ as the
number density of infected persons at each $s_{n}$ and $\beta_{n}\sim h\beta(s_{n})$ for $n\neq0,N$
and $\beta_{n}\sim h\beta(s_{n})/2$ for $n=0,N$. As in section \ref{subsec:SIkR}, for best results
the value of $\beta_{n}$ should be re-scaled so that $\sum_{n=1}^{N}\beta_{n}=1$. The time-step $k=0,1,2,...$
will be indicated as a superscript in parentheticals, e.g. $S^{(k)}$ is the susceptible population at
time $t^{(k)}$. In the language of a predictor/corrector method, intermediate predictions will be further
denoted by the letter $p$ inside the parenthetical, e.g. $S^{(k+1,p)}$.

Beginning from time $t^{(k)}$, we can calculate the state of the system at time $t^{(k+1)}=t^{(k)}+h$
as follows. We begin with an explicit prediction step:

\begin{equation}
\Delta S^{(k)}=h\Bigg[-R_{0}S^{(k)}\sum_{n=1}^{N}\beta_{n}I_{n}^{(k)}\Bigg]
\end{equation}

\begin{equation}
S^{(k+1,p)}=S^{(k)}+\Delta S^{(k)}
\end{equation}

\begin{equation}
I_{1}^{(k+1,p)}=-\Delta S^{(k)}/h
\end{equation}

\begin{equation}
I_{n>1}^{(k+1,p)}=I_{n-1}^{(k)}
\end{equation}
Then, we proceed to the correction:

\begin{equation}
\Delta S^{(k+1,p)}=h\Bigg[-R_{0}S^{(k+1,p)}\sum_{n=1}^{N}\beta_{n}I_{n}^{(k+1,p)}\Bigg]
\end{equation}

\begin{equation}
S^{(k+1)}=S^{(k)}+\frac{1}{2}\Big(\Delta S^{(k)}+\Delta S^{(k+1,p)}\Big)
\end{equation}

\begin{equation}
I_{1}^{(k+1)}=-\Delta S^{(k+1,p)}/h
\end{equation}

\begin{equation}
I_{n>1}^{(k+1)}=I_{n}^{(k+1,p)}
\end{equation}

\begin{equation}
R^{(k+1)}=R^{(k)}+\frac{h}{2}\Big(I_{N}^{(k)}+I_{N}^{(k+1)}\Big)\label{eq:R_PC}
\end{equation}
Once again, equation \ref{eq:R_PC} applies to the special case where a person recovers only after aging
beyond the maximum infectious period, $\phi_{R}(s)=\delta(s-1)$. For more realistic versions of $\phi_{R}(s)$,
equation \ref{eq:R_PC} becomes:

\begin{equation}
R^{(k+1)}=R^{(k)}+\frac{h}{2}\sum_{n=1}^{N}\phi_{R,n}\Big(I_{N}^{(k)}+I_{N}^{(k+1)}\Big)
\end{equation}

\begin{equation}
\phi_{R,n}\sim\begin{cases}
h\phi_{R}(s_{n})/2 & \text{if }n=1\text{ or }n=N\\
h\phi_{R}(s_{n}) & \text{otherwise}
\end{cases}
\end{equation}

\begin{equation}
\sum_{n=1}^{N}\phi_{R,n}=1
\end{equation}

This method has second order accuracy $O(h^{2})$, so compared to the SIkR model one needs only a few
stages to obtain good results. The principle weakness of this method is that time-stepping is inflexible
-- each time step must advance the solution forward by a time of exactly $h$. Note also that for numerical
stability, the time-step $h$ must be chosen to be sufficiently small (e.g. $h(R_{0}-1)<1$).

\subsection{\label{subsec:Galerkin}Expansion in Legendre Polynomials (Galerkin)}

For spectral accuracy in $s$ and flexible time-stepping, we provide a discretization based on a Galerkin
approximation. On the interval $s\in[-1,1]$, we can write $I(t,s)$ as weighted sum of Legendre polynomials:

\begin{equation}
I(t,s)=\sum_{n=0}^{\infty}I_{n}(t)P_{n}(s)
\end{equation}
Any set of orthogonal basis functions could be used for this purpose, but we have found that Legendre
polynomials work particularly well in this application. If we truncate the sum after $m+1$ terms and
insert into the governing equations, we obtain a closed system of equations for $S,I_{0},I_{1},I_{2},\ldots,I_{m},$
and $R$. This technique is known as a Galerkin approximation.

The population of susceptibles evolves by:

\begin{equation}
\frac{dS}{dt}=-R_{0}S\sum_{n=0}^{m}a_{n}I_{n}
\end{equation}
where the coefficients $a_{n}$ are:

\begin{equation}
a_{n}=\int_{-1}^{1}ds\beta(s)P_{n}(s)
\end{equation}
For $n<m$, we obtain evolution equations for the coefficients $I_{n}$:

\begin{equation}
\frac{dI_{n}}{dt}+\sum_{k=0}^{m}b_{nk}I_{k}=0
\end{equation}

\begin{equation}
b_{nk}=\Bigg[\frac{2n+1}{2}\Bigg]\int_{-1}^{1}dsP_{n}(s)P'_{k}(s)=\begin{cases}
2n+1 & \text{if }n+k\text{ is odd and }k>n\\
0 & \text{otherwise}
\end{cases}
\end{equation}

The highest order polynomial coefficient $I_{m}$ is constrained by the boundary condition at $s=-1$
(c.f. equation \ref{eq:ndim_bc}):

\begin{equation}
\sum_{n=0}^{m}I_{n}(-1)^{n}=R_{0}S\sum_{n=0}^{m}a_{n}I_{n}\label{eq:Galerkin bc}
\end{equation}
In practice, this constraint can sometimes be used to eliminate the highest order Legendre polynomial
coefficient, $I_{m}$, as an independent variable.

\begin{equation}
I_{m}=[(-1)^{m}-R_{0}Sa_{m}]^{-1}\Bigg[SR_{0}\sum_{n=0}^{m-1}a_{n}I_{n}-\sum_{n=0}^{m-1}I_{n}(-1)^{n}\Bigg]\label{eq:Galerking_BC_explicit}
\end{equation}
However, this trick of eliminating $I_{m}$ works best when the reproduction number $R_{0}$ (or more
generally, the contact matrix in an age-structured model) is not varying in time (c.f. section \ref{sec:Galerkin_problems}).
Implicit algebraic constraints can be costly during numerical integration, so the predictor/corrector
method may be preferred in these cases.

Finally, we consider the growth dynamics for a recovered class:

\begin{equation}
\frac{dR}{dt}=\sum_{n=0}^{m}\phi_{R,n}I_{n}
\end{equation}

\begin{equation}
\phi_{R,n}=\int_{-1}^{1}ds\phi_{R}(s)P_{n}(s)
\end{equation}

The Galerkin discretization does offer spectral resolution in $s$ and flexible time-stepping -- both
of which are improvements over the second-order accuracy and inflexible time-stepping of the predictor-corrector
method. However, as is often the case with spectral methods, the improvements in accuracy come with costs
regarding the robustness of the method. These costs will be discussed in more detail in section \ref{sec:Galerkin_problems}.

\section{\label{sec:Sample-Results:-Simple}Sample Results: Simple Calculations}

In this section, we will provide sample calculations for each of the discretization schemes given in
the preceding sections. In each case, we will solve for the growth dynamics of a system with $R_{0}=2$
and $\beta(s)\sim(1+s)(1-s)^{4}$. We also consider non-trivial recovery dynamics, $\phi_{R}(s)=(1+s)^{4}(1-s)$.
This choice of input parameters is intended for proof-of-concept calculations only and does not purport
to represent the dynamics of any particular disease. For an initial condition, we seed with a small infected
population distributed in $s$ to follow the fastest growing linear mode.

\subsection{Linear Growth}

To find the fastest growing linear mode, we return to the PDE formulation of the model. From equation
\ref{eq:ndim_I} we find:

\begin{equation}
I(t,s)=I_{0}\exp(\lambda(t-s))
\end{equation}
with growth rate $\lambda$ and density of recent infecteds $I_{0}\ll1$. From the boundary condition
at $s=-1$ we also have:

\begin{equation}
I(t,-1)=I_{0}\exp(\lambda(t+1))
\end{equation}
This gives an implicit relationship between the growth rate $\lambda$ and the reproduction number $R_{0}$
and $\beta(s)$ that can be solved numerically:

\begin{equation}
\frac{1}{R_{0}S}=\int_{-1}^{1}\beta(s)\exp(-\lambda(s+1))ds\label{eq:eigenvalue_equation}
\end{equation}

Thus, we see that the growth rate $\lambda$ (in units of $2/T$) is not uniquely determined by the reproduction
number $R_{0}$ but also depends on how the infectiousness of a disease rises and falls over time within
the infectious period. Equivalent results have been previously reported for TSI models \cite{key-13}.

In this section, all simulations begin from an initial condition that approximates the fastest growing
linear mode, $I(0,s)=10^{-3}\exp(-\lambda s)$. Analysis of linear growth dynamics for an age-structured
version of the TSI model are provided in the appendix to the present report.

\subsection{\label{subsec:Nonlinear-Dynamics}Nonlinear Dynamics}

\begin{figure}[H]
\begin{centering}
\includegraphics[scale=0.6]{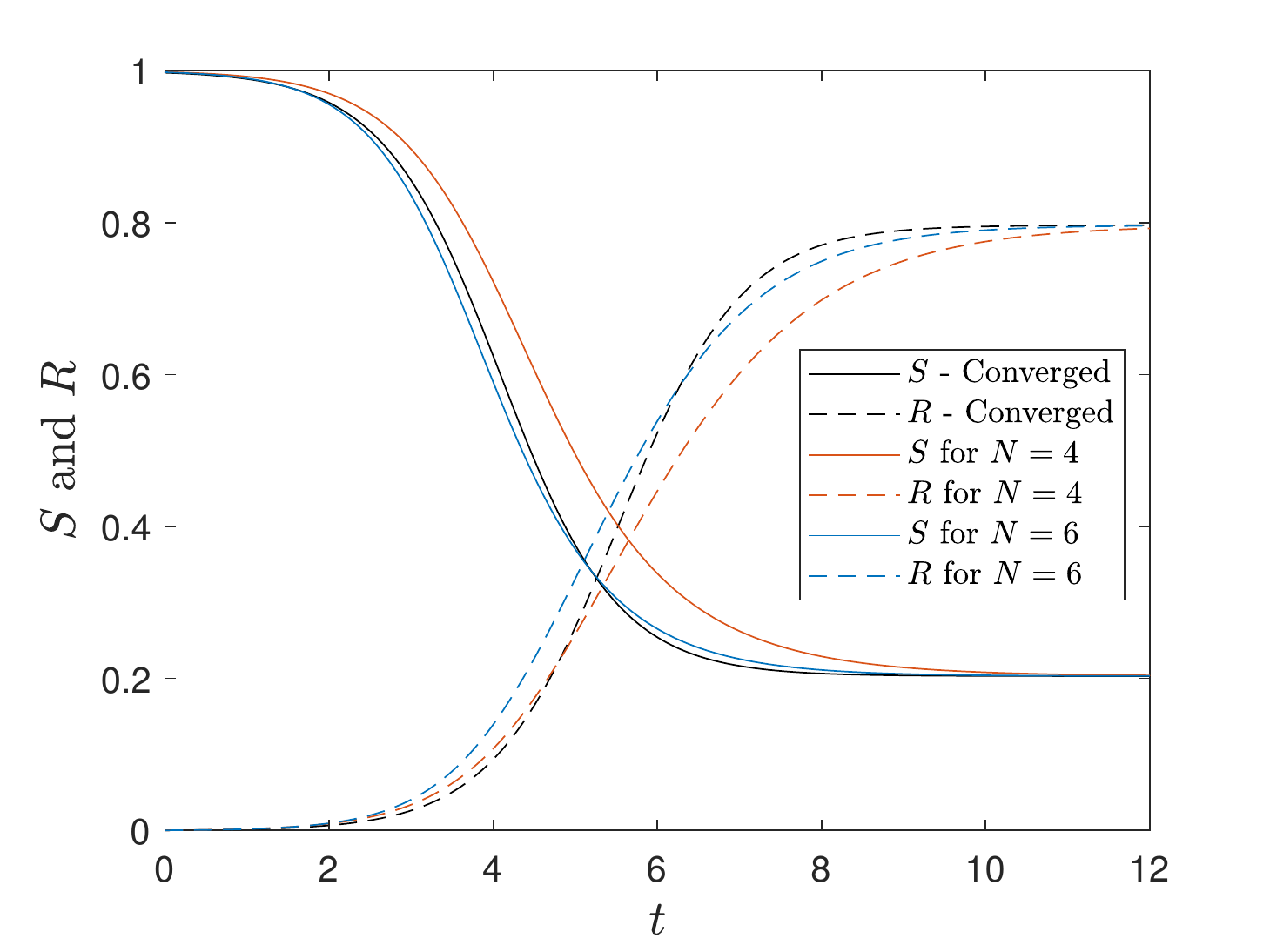}
\par\end{centering}
\caption{\label{fig:SIkR_example}Showing convergence of SIkR calculations with $R_{0}=2$, $\beta(s)\sim(1+s)(1-s)^{4}$,
and $\phi_{R}(s)=(1+s)^{4}(1-s)$.}
\end{figure}

First, we consider the SIkR discretization, per section \ref{subsec:SIkR}. In Figure \ref{fig:SIkR_example},
we see that there is noticeable error for small values of $N\leq6$ and trends for convergence are not
evident between $N=4$ and $N=6$. The convergence of the SIkR discretization is just $O(h)$, so it
is not surprising that the model performs so poorly here.

\begin{figure}[H]
\begin{centering}
\includegraphics[scale=0.6]{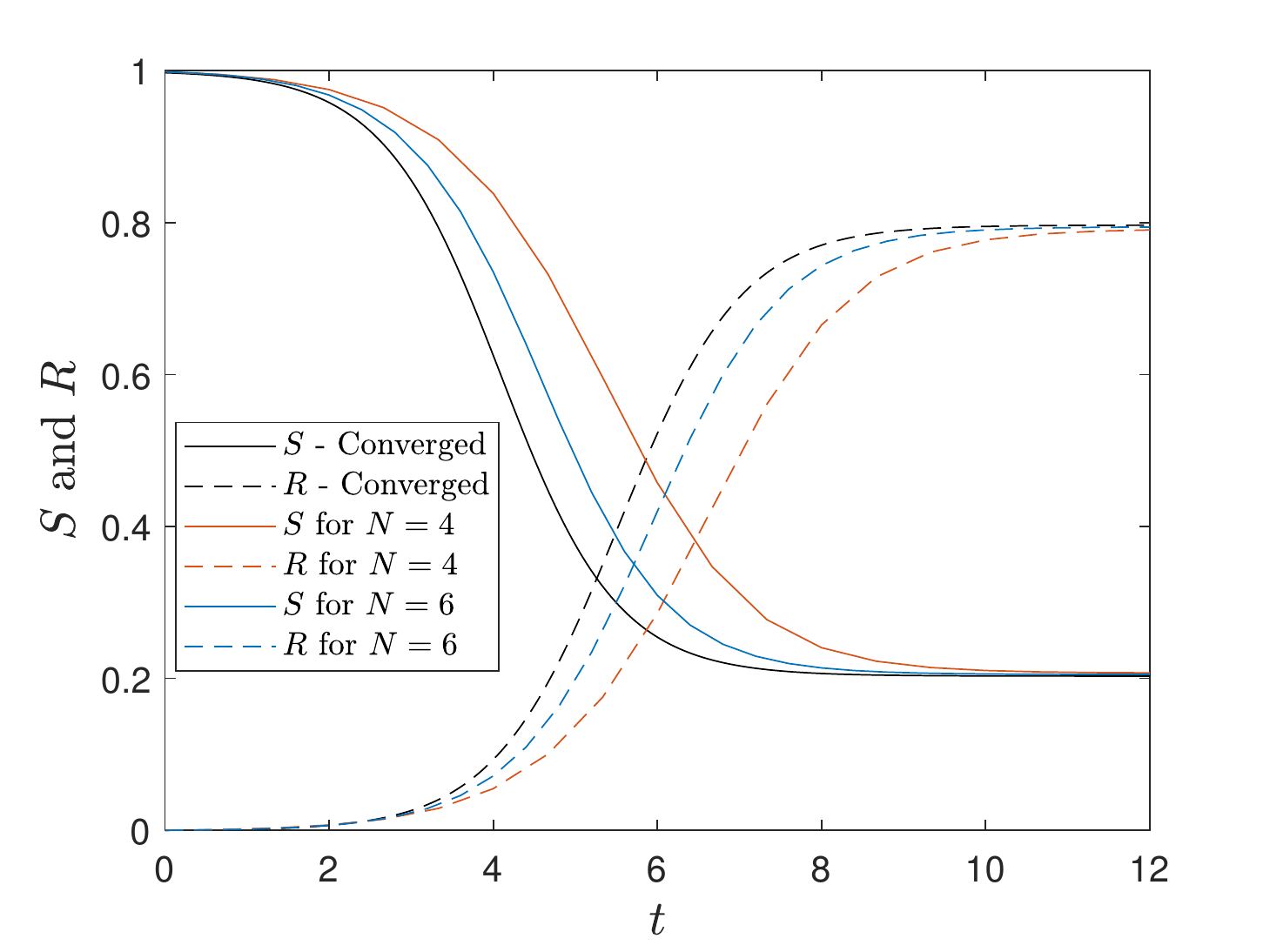}
\par\end{centering}
\caption{\label{fig:PC_example}Showing convergence of predictor/corrector calculations with $R_{0}=2$, $\beta(s)\sim(1+s)(1-s)^{4}$,
and $\phi_{R}(s)=(1+s)^{4}(1-s)$.}
\end{figure}

Next, we consider the predictor/corrector (PC) discretization, per section \ref{subsec:PC}. In Figure
\ref{fig:PC_example}, we see that the absolute error at small $N$ happens to be larger in this case,
but the trends for convergence are clear, and the PC method will vastly out-perform the SIkR method for
slightly larger values of $N$. Note also that the PC method takes large time steps, so the cost to implement
a larger value of $N$ is still very good by comparison.

\begin{figure}[H]
\begin{centering}
\includegraphics[scale=0.6]{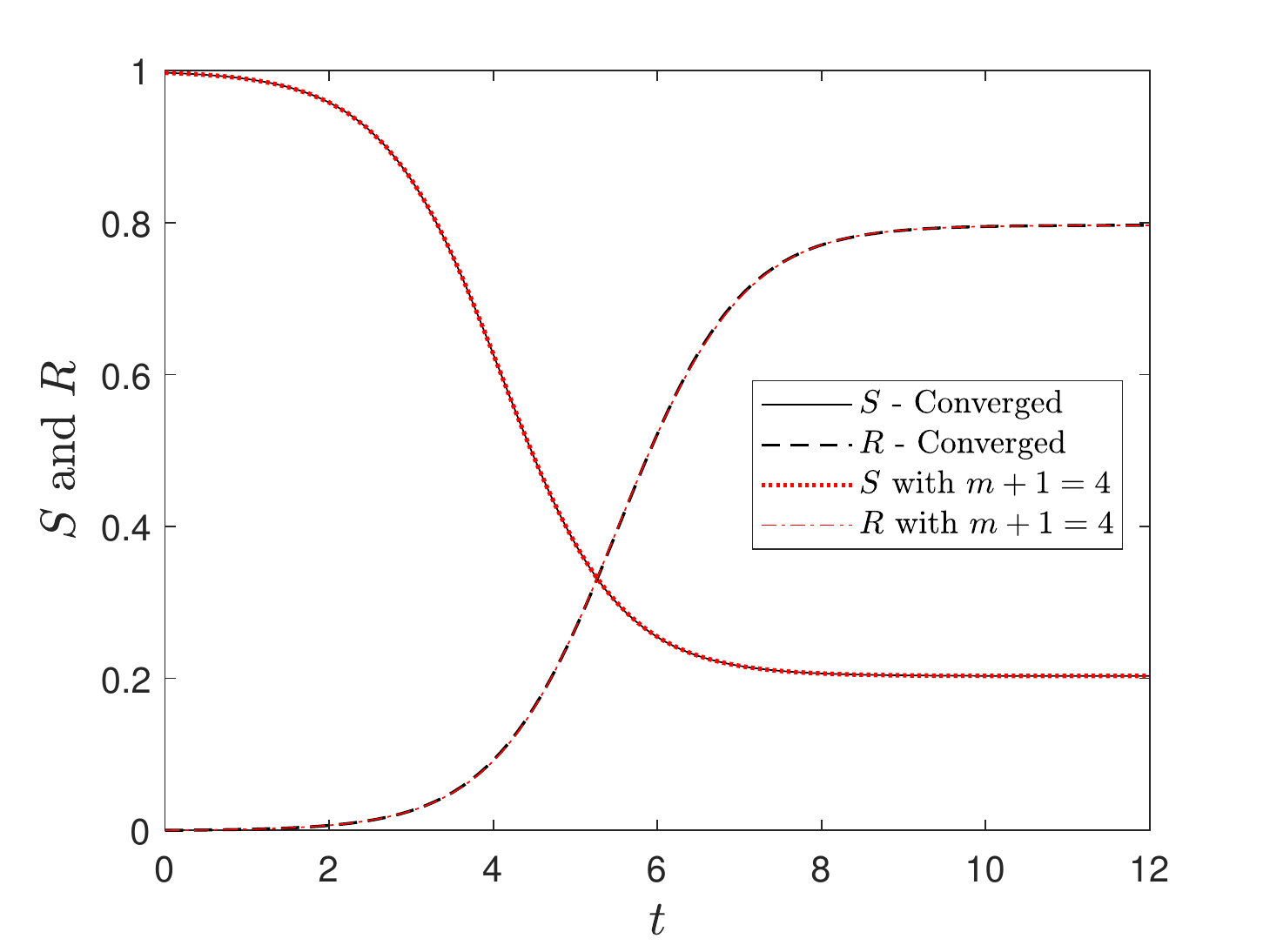}
\par\end{centering}
\caption{\label{fig:Galerkin_example}Showing convergence of calculations with the Galerkin model for $R_{0}=2$,
$\beta(s)\sim(1+s)(1-s)^{4}$, and $\phi_{R}(s)=(1+s)^{4}(1-s)$.}
\end{figure}

Finally, we consider the Galerkin discretization, per section \ref{subsec:Galerkin}. In Figure \ref{fig:Galerkin_example},
we see that the calculations are essentially converged to the thickness of the line shown with just $m+1=4$
modes (corresponding to $m=3$ independent modes using equation \ref{eq:Galerking_BC_explicit} to eliminate
the highest order Legendre polynomial coefficient).

In this specific application, it is clear that the Galerkin method vastly out-performs the predictor/corrector
and SIkR discretization strategies. This should not be too surprising, since the Galerkin discretization
provides spectral accuracy in $s$. However, as mentioned in section \ref{subsec:Galerkin}, models with
spectral accuracy can be less robust in certain applications. In the following section, we discuss common
problems that one might encounter with the Galerkin method.

\section{\label{sec:Galerkin_problems}When to avoid Galerkin}

Considering the Galerkin method, the dramatic improvements in accuracy come at a cost in the robustness
of the method. In this section, we will discuss problems that the Galerkin method has with (1) abrupt
changes in $R_{0}$, (2) smooth changes in $R_{0}$, and (3) arbitrarily chosen initial conditions.

For the first two, we must explain what it means to think about abrupt/smooth changes in $R_{0}$. To
this point, we have described the reproduction number as a dimensionless constant, but in this section
we will allow it to be a function of time, $R_{0}=R_{0}(t)$, as though the underlying dynamics of transmission
were also subject to some time-dependent modulation $\beta(t,s)=u(t)\beta(s)$. Physically, this construction
might be useful for describing changes in non-pharmaceutical interventions (e.g. social distancing, lockdowns,
etc.) or the inherent seasonality of an epidemic.

\begin{figure}[H]
\begin{centering}
\includegraphics[scale=0.6]{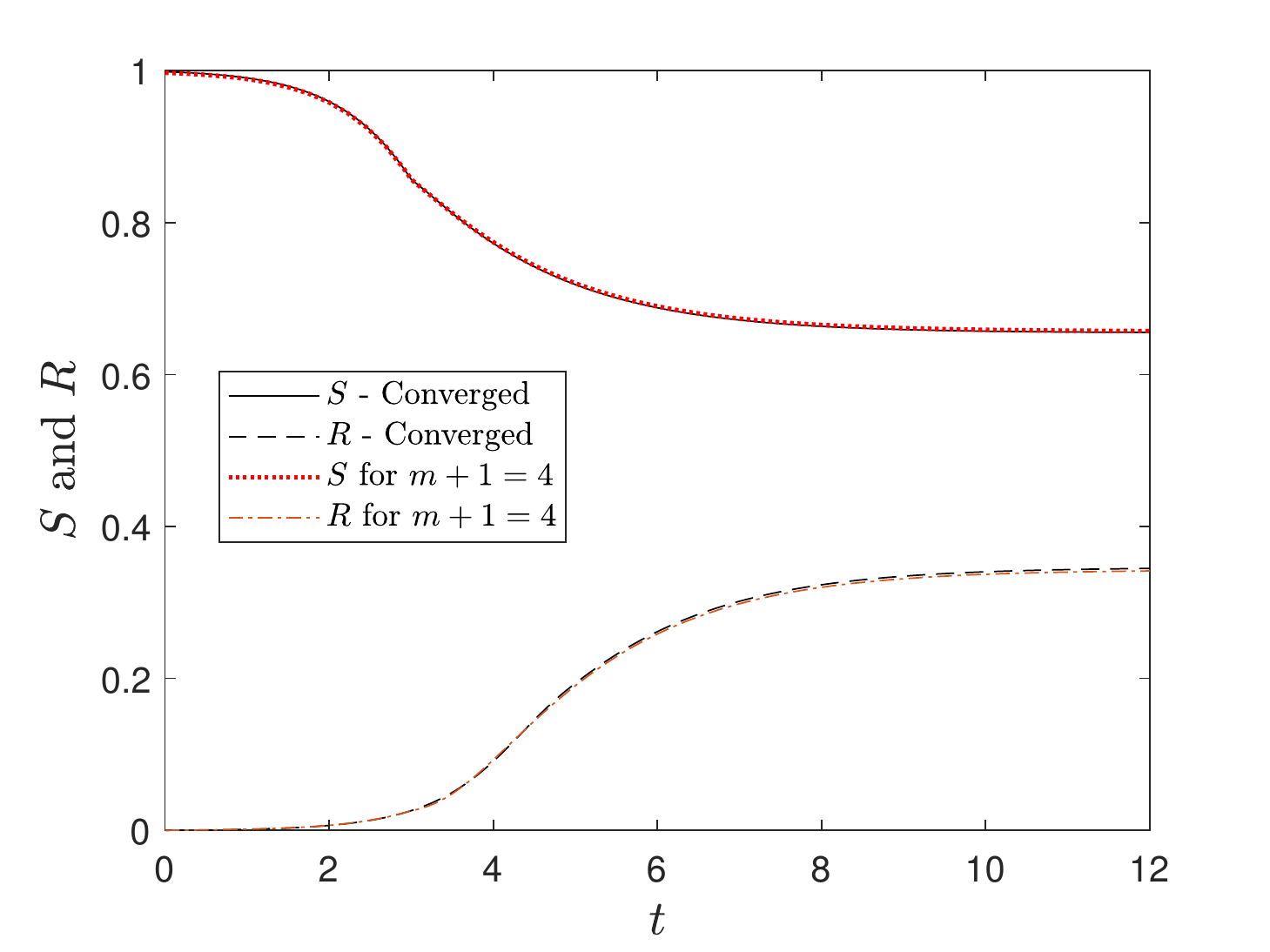}
\par\end{centering}
\caption{\label{fig:Galerkin_vs_PC_lockdown} Comparing converged predictions of the PC method with predictions
from the Galerkin method, $m+1=4$ when there are abrupt changes in the reproduction number. Here, we
consider $R_{0}(t<3)=2,$$R_{0}(t>3)=1,$ $\beta(s)\sim(1+s)(1-s)^{4}$, and $\phi_{R}(s)=(1+s)^{4}(1-s)$.}
\end{figure}

If there are step-wise changes or 'kinks' in $R_{0}(t)$ at some time $t_{0}$, then the distribution
of infecteds $I(t,s)$ within the infectious range $s\in[-1,1]$ will also be non-smooth for all times
$t\in[t_{0},t_{0}+2]$. Spectral methods (including those based on Legendre Polynomials) are less accurate
when representing non-smooth functions, so the Galerkin discretization will sometimes (but not always!)
perform poorly following non-smooth dynamics in $R_{0}(t)$. As an example, we replicate the calculations
of the preceding section but impose $R_{0}(t>3)=1$. In Figure \ref{fig:Galerkin_vs_PC_lockdown}, we
compare the converged solution from the PC method with the Galerkin result for $m+1=4$. In this case,
the Galerkin method is not as good as it was in Figure \ref{fig:Galerkin_example}, but overall it still
appears to match the converged solution very well.

The quality of agreement found in Figure \ref{fig:Galerkin_vs_PC_lockdown} is perhaps partially attributed
to the smooth choice of functions $\beta(s)$ and $\phi_{R}(s)$. If we instead choose $\phi_{R}(s)=\delta(s-1)$,
we find that shortly after the change in $R_{0}$, the recovered population temporarily decreases --
a result that is clearly non-physical.

\begin{figure}[H]
\begin{centering}
\includegraphics[scale=0.6]{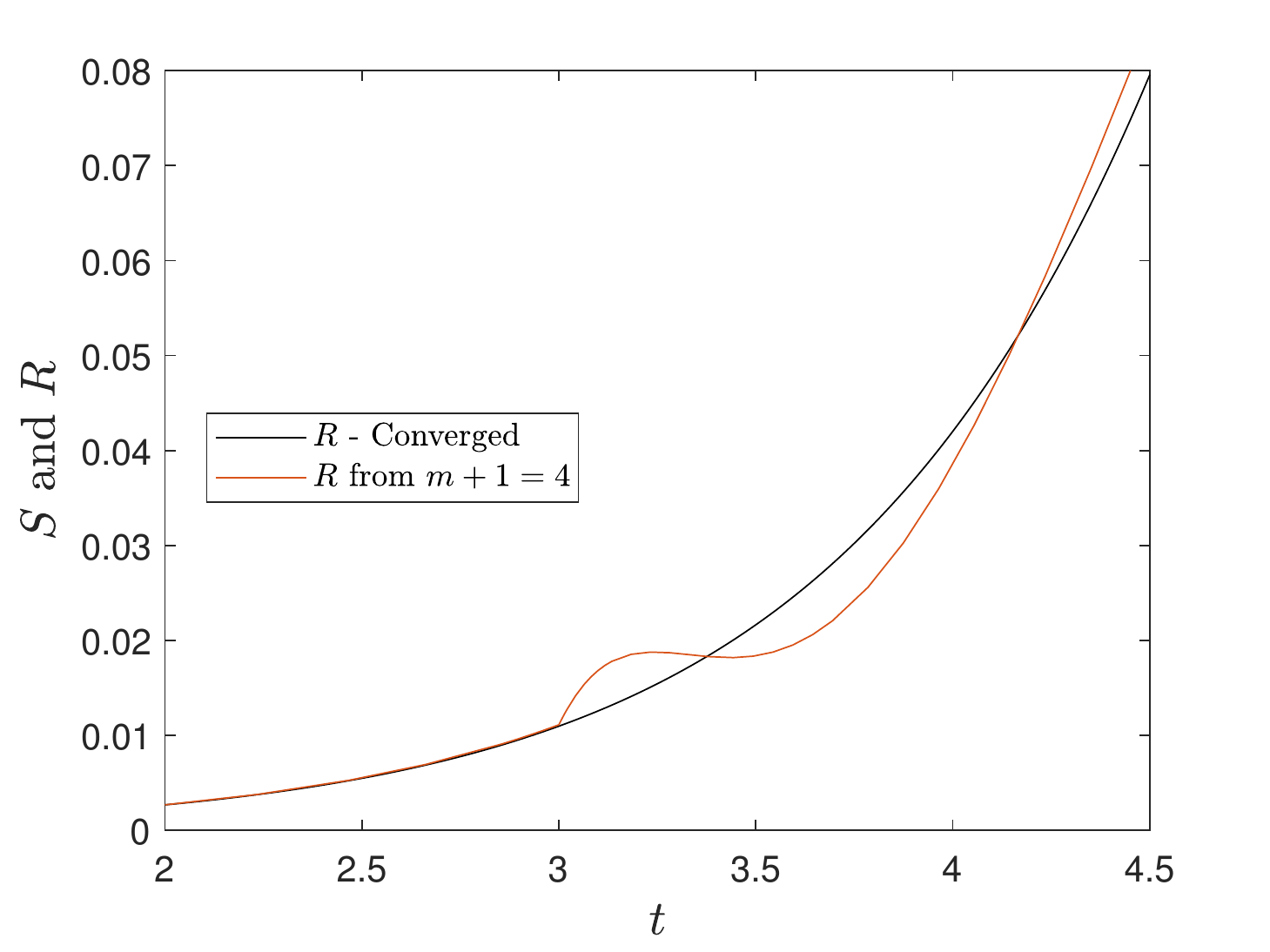}
\par\end{centering}
\caption{\label{fig:Galerkin_vs_PC_lockdown-1} Comparing converged predictions of the PC method with predictions
from the Galerkin method, $m+1=4$ when there are abrupt changes in the reproduction number and a non-smooth
$\phi_{R}(s)$. Here, we consider $R_{0}(t<3)=2,$$R_{0}(t>3)=1,$ $\beta(s)\sim(1+s)(1-s)^{4}$, and
$\phi_{R}(s)=\delta(s-1)$.}
\end{figure}

As a matter of best practice, one may want to consider using other methods whenever $R_{0}(t)$ is non-smooth
-- for example, it might be better to use the predictor-corrector method for the short interval $t\in[t_{0},t_{0}+2]$.

A similar problem also emerges whenever one attempts to apply an arbitrary choice of initial condition.
The initial condition contains a history of infections going back to time $t=-2$. If the choice of initial
condition is not consistent with the same constant value of $R_{0}$ that is applied for $t>0$, the
distribution of infecteds $I(t,s)$ within the infectious range $s\in[-1,1]$ will be non-smooth until
time $t=2$. It is for this reason that the starting condition for all the simulations in section \ref{subsec:Nonlinear-Dynamics}
was chosen to be aligned with the fastest growing linear mode. As a matter of best practice, when applying
an arbitrary initial condition we suggest using the PC method to integrate to time $t=2$.

\begin{figure}[H]
\centering{}\includegraphics[scale=0.6]{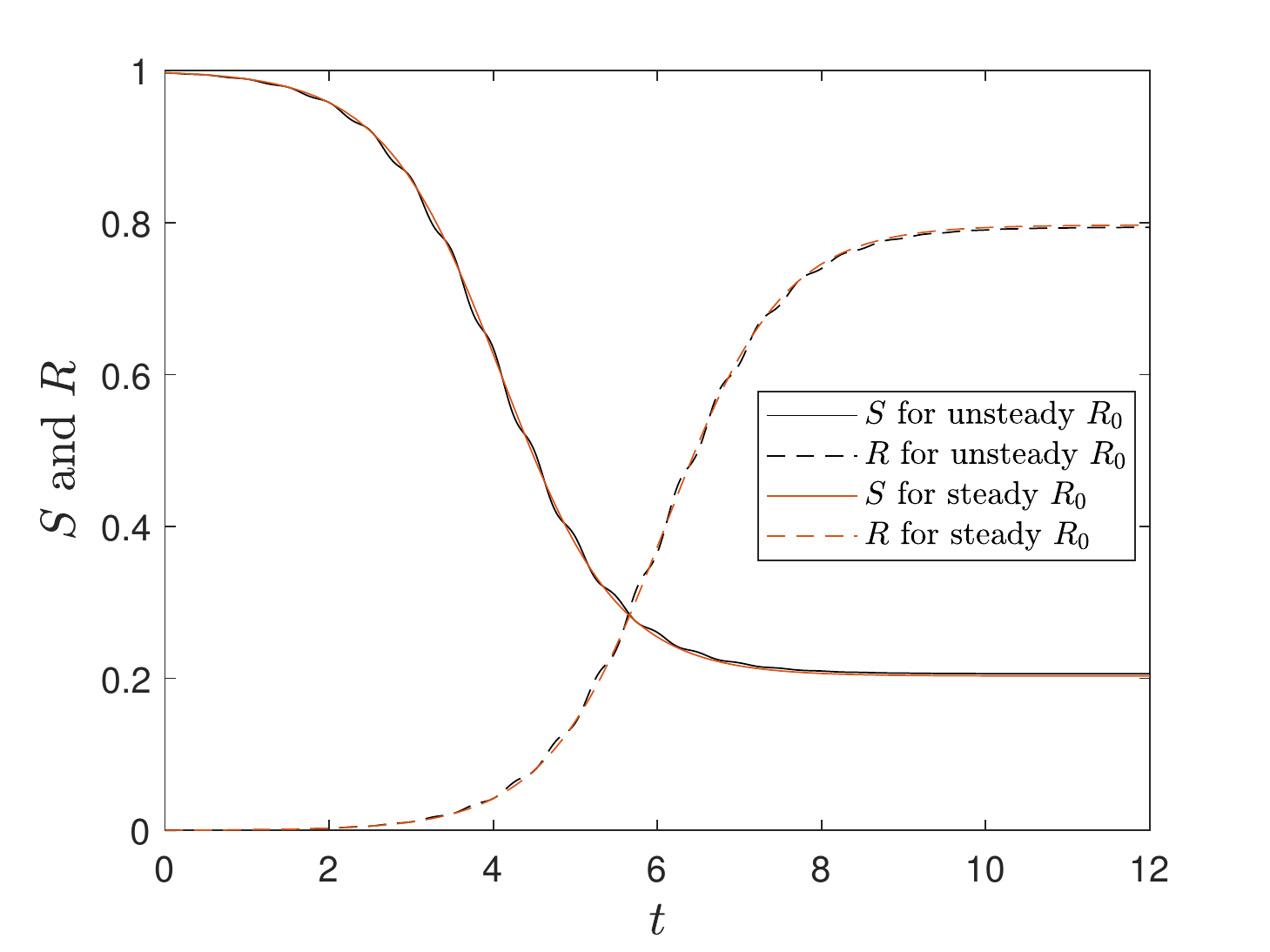}\caption{\label{fig:steady vs unsteady R0=00003D2}Comparing converged predictions for the TSI model with constant
reproduction number $R_{0}=2.5$ (solid lines) with a time-varying reproduction number $R_{0}=2.5(1+.5\sin(2\pi t))$.
The large oscillations in $R_{0}$ only translate to a small oscillation about the solution obtained
with a constant reproduction number of the same mean value $R_{0}=2.5$.}
\end{figure}

However, even with a good choice of initial conditions and a smoothly varying $R_{0}(t)$, there are
potential limitations to the Galerkin method. In any TSI method (including PC and SIkR), one must always
be sure that the numerical method has sufficient resolution to capture the unsteady behavior in $R_{0}(t)$:
when $R_{0}(t)$ is varying on timescales that are fast compared to the infection period, a higher number
of modes will be needed to capture that variation. For example, if the reproduction number is changing
hour-by-hour and the infectious period is several weeks, one will need hundreds or thousands of modes
to capture all the effects of a time-varying reproduction number. Fortunately, whenever the reproduction
number varies on such fast time-scales, one can often use a smoothed out approximation (e.g. moving average)
to obtain very similar results in the trajectories for $S$ and $R$ with a much smaller number of modes.
In Figure \ref{fig:steady vs unsteady R0=00003D2}, we compare converged predictions for $R_{0}=2$ and
$R_{0}=2+\sin(2\pi t)$ with the same values of $\beta(s)$ and $\phi_{R}(s)=\delta(s-1)$. The unsteady
reproduction number requires substantially more modes ($m+1\approx16$) for convergence, and unrealistic
results (c.f. Figure \ref{fig:Galerkin_vs_PC_lockdown-1}) are seen when too few modes are present.

Finally, whenever the reproduction number is varying in time -- on any timescale -- we have empirically
found that there are minor problems with finding an approximation to $I(t,s)$ that approximately preserves
the property $I(t,s>t-1)=I(t-(s+1),-1)$ whenever the boundary condition \ref{eq:Galerkin bc} is handled
explicitly per equation \ref{eq:Galerking_BC_explicit}. Fortunately, we have not yet found an example
in which this particular difficulty in describing $I(t,s)$ leads to notable errors in the trajectories
for $S$ and $R$. Whatever the underlying issue here may be, we have found that it is absolved when
the boundary condition \ref{eq:Galerkin bc} is handled implicitly, but this will often make implementation
of the Galerkin method much more computationally expensive.

\section{\label{sec:Sample-Results:-Optimal}Sample Results: Optimal Control}

A principal motivation of the present report is to develop numerical methods for TSI models that are
computationally tractable for solving complex problems in the fight against infectious diseases. The
example calculations given in section \ref{subsec:Nonlinear-Dynamics} do affirm the computational efficacy
of our methods, but they do not represent a complex calculation that could not be reasonably attained with
pre-existing (and less efficient) methods. 
Therefore, in this section we will provide proof-of-concept results for
optimal control caclulations -- calculations that (to our knowledge) have not yet been reported for
a TSI model.

Here, we use the same model parameters as in section \ref{subsec:Nonlinear-Dynamics}, but we consider
a time-dependent $R_{0}(t)=u(t)R_{0}$, where $u(t)$ is a control function that modulates the effective
reproduction number. This control function could describe non-pharmaceutical interventions such as social
distancing and lockdowns to prevent the transmission of a disease. For simplicity, we choose to describe
the control function $u(t)$ as a piece-wise linear interpolation between a set of $N_{u}$ interpolation
points. Since this makes $R_{0}(t)$ non-smooth, we employ the predictor-corrector method of section
\ref{subsec:PC} to evolve the epidemic forward in time. The position and value of the interpolation
points are locally optimized using the COBYLA algorithm in the optimization package 'NLopt'. We also
use the epidemic modelling package 'pyross' to solve the epidemic model using the predictor-corrector
method of section \ref{subsec:PC} with $N=16$.

Optimal control calculations require a cost function, which we represent as the sum of a social distancing
cost and the cost of infections evaluated on an interval $t\in[0,T_{f}]$, with $T_{f}=35$. The cost
of infections is given by the total number of infections $1-S$ at the end time, and we assume the cost
of modulating the reproduction number at any time scales as $(1-u)^{2}S$. We use the parameter $\Omega$
as a constant of proportionality between these two costs so that the total cost $C$ is given by:

\begin{equation}
C=1-S(T_{f})+\Omega\int_{0}^{T_{f}}dt(1-u(t))^{2}S(t)
\end{equation}

Large values of $\Omega$ will represent diseases that are (a) not very deadly and/or (b) extremely costly to contain (e.g.
the common cold), whereas small values of $\Omega$ describe epidemics that are more deadly and/or less
costly to contain (e.g. Ebola or MERS).  Note that our analysis here has not considered the possibility of re-infection - for diseases with short-lived immunity, the conversation on optimal control can be completely different.

\begin{figure}[H]
\begin{centering}
\includegraphics[scale=0.6]{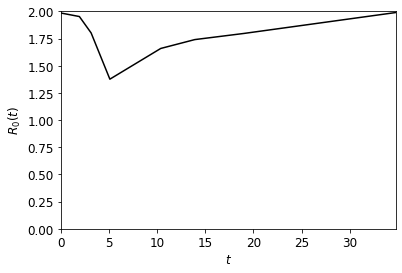}
\par\end{centering}
\caption{\label{fig:R0_optimal_control}Optimal control solution ($\Omega=0.3$) for $R_{0}(t)$ in an epidemic
with base reproduction number $R_{0}=2$, $\beta(s)\sim(1+s)(1-s)^{4},$and $\phi_{R}(s)\sim(1+s)^{4}(1-s)$}
\end{figure}

\begin{figure}[H]
\begin{centering}
\includegraphics[scale=0.6]{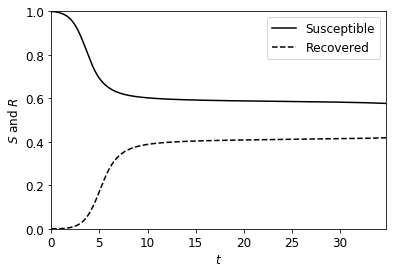}
\par\end{centering}
\caption{\label{fig:R0_optimal_control-1}Predictions for $S(t)$ and $R(t)$ under the optimal control solution
($\Omega=0.3$) for $R_{0}(t)$ in an epidemic with base reproduction number $R_{0}=2$, $\beta(s)\sim(1+s)(1-s)^{4},$and
$\phi_{R}(s)\sim(1+s)^{4}(1-s)$.}
\end{figure}

Naievely, one might think that given a TSI model with a maximum infectious period (and no possibility of re-infection), the best control solution
will always be to eradicate the disease with a full lockdown lasting one infection period,
giving a cost $C=2\Omega$. However, this is not practical for mild diseases, $\Omega \gg 1$, which are
not deadly enough to warrant such drastic measures. For sufficiently large values of $\Omega$ one will
always find that that the best option is to let the disease run its course with targeted interventions
that guide the system towards a less costly 'herd immunity' state. As an example, we present results
for optimal control with $\Omega=0.3$, for which a solution with $C=0.54<2\Omega$ has been found. The
time-dependent $R_{0}(t)$ is shown in Figure \ref{fig:R0_optimal_control}, and the resulting evolution
of $S$ and $R$ are shown in Figure \ref{fig:R0_optimal_control-1}. Note that the interventions mostly
occur after the epidemic has already peaked (the inflection point in $S(t)$). This is because for large
values of $\Omega$, containing the disease is futile -- the focus is not on preventing the epidemic
but on finding a less costly path to herd immunity.

\section{Discussion and Conclusions}

Epidemic models are useful tools in the fight against infectious diseases, but they are often forced
to strike a balance between accuracy and complexity. Infection-age or time since infection (TSI) models
allow for a more realistic description of disease dynamics, but a percieved burden of computational complexity
has here-to limited their use in many applications.  In the present work,
we have shown that one can retain a full TSI description of disease transmission with only a small increase in the
computational costs, especially if one employs our Galerkin discretization strategy. We have
discussed the advantages (accuracy with low modes) and disadvantages (robustness to non-smooth dynamics)
of the Galerkin approximation, and also provided proof-of-concept calculations for 'optimal control'
using a 'predictor/corrector' scheme. Besides providing more efficient numerical methods, we have also devised a
'filtering' technique that eliminates the need to assign different disease dynamics to different sub-populations of
the infected population.  Finally, the governing equations for an age-structured generalization of
the model are given in the appendix. At this point, it is our view that TSI models can now be employed
for a wide range of applications in which they were previously not considered feasible. The tools and
methods described in the present report (including the age-structured version) are all freely available
through the software package pyross.

\section{Acknowledgments}

This work was undertaken as a contribution to the Rapid Assistance in Modelling the Pandemic (RAMP) 
initiative, coordinated by the Royal Society. This work was funded by the European Research 
Council, under the Horizon 2020 Programme, ERC grant 740269.  The authors would especially like to thank 
Simon Frost and Robert Jack for their valuable feedback on this project, and also to Rajesh Singh for 
his help incorporating the methods outlined in this report into the epidemic modelling framework of pyross.

\appendix

\section{Class-structured TSI model}

\subsection{Dimensional Governing Equations}

In many cases, it is important to resolve predictions for how an epidemic will spread through different
sub-populations (e.g. age, occupation, socioeconomic status, etc.). If there are $C$ classes in the
population, we can denote population that is susceptible and assigned to class $i=1,2,\ldots,M$ as $S_{i}$.
Likewise, in this appendix the population that is assigned to class $i$ and was infected during the
narrow interval of$s\in[s,s+\delta s]$ in the past is $I_{i}(t,s)\delta s$, and the total population
assigned to class $i$ is $N_{i}$. The mean frequency of contact between members of class $i$ and $j$
is given by the quantity $C_{ij}/N_{j}$, where $C_{ij}$is referred to as the 'contact matrix'. In the
most general case, we allow each class $i$ to have a different value of $\beta_{i}(s)$. The full PDE
version \footnote{The age compartments arise via a discretization of a more general PDE model in which age is treated as
a proper continuous variable. Here, we have effectively employed a midpoint discretization of the full
age structure and also neglected the usual population dynamics (birth, aging, and natural death) on the
basis that the epidemic time is typically much much shorter than the typical lifespan.} of the governing equations are given by:

\begin{equation}
\frac{dS_{i}}{dt}=-S_{i}\sum_{j=1}^{M}\frac{C_{ij}}{N_{j}}\int_{0}^{T}ds\beta_{j}(s)I_{j}(t,s)
\end{equation}

\begin{equation}
\frac{\partial}{\partial t}I_{i}(t,s)+\frac{\partial}{\partial s}I_{i}(t,s)=0
\end{equation}

\begin{equation}
I_{i}(t,0)=S_{i}\sum_{j=1}^{M}\frac{C_{ij}}{N_{j}}\int_{0}^{T}ds\beta_{j}(s)I_{j}(t,s)
\end{equation}

Sub-classes of the infected population (e.g. infectious, recovered, hosptialized, symptomatic, asymptomatic,
etc.) can be computed in the same manner as before. For example, if we assume that all infecteds are
classified as recovered once their infection ages beyond time $T$, then the fraction of the population
that is recovered and assigned to sub-class $i$ evolves as:

\begin{equation}
\frac{dR_{i}}{dt}=I_{i}(t,T)
\end{equation}

\subsection{Dimensionless Governing Equations}

As before, we prefer to work with the governing equations in a dimensionless form. We rescale $S_{i}$
and $R_{i}$ by the total population $N=\sum_{i}N_{i}$, such that $\tilde{S_{i}}=S_{i}/N$ and $\tilde{R_{i}}=R_{i}/N$.
Using a characteristic time $T_{C}=T/2$, we rescale time as $\tilde{t}=t/T_{C}$ and number density
of infecteds as $\tilde{I_{i}}=I_{i}C/N$. The time since infection is rescaled to the variable $\tilde{s}=s/T_{C}-1$
so that the range $s\in[0,T]$ is mapped to $\tilde{s}\in[-1,1]$. Finally, we define a class-specific
reproduction number

\begin{equation}
R_{0,i}=\sum_{j=1}^{M}\frac{C_{ij}}{N_{j}}\int_{0}^{T}ds\beta_{j}(s)
\end{equation}
and a dimensionless matrix of rate constants that subsumes information about the contact structure:

\begin{equation}
\tilde{\beta}_{ij}(s)=\frac{T_{C}}{R_{0,i}}\frac{C_{ij}}{N_{j}}\beta_{j}(s)
\end{equation}
Note that the matrix $\tilde{\beta}_{ij}$ has the property:

\begin{equation}
\sum_{j=1}^{M}\int_{-1}^{1}d\tilde{s}\tilde{\beta}_{ij}(\tilde{s})=1
\end{equation}
For compactness of notation, we will suppress tildes in our dimensionless variables -- from this point
forward, all variables are dimensionless unless otherwise stated. The dimensionless governing equations
for the age-structured TSI model are given by:

\begin{equation}
\frac{dS_{i}}{dt}=-R_{0,i}S_{i}\sum_{j=1}^{M}\int_{-1}^{1}ds\beta_{ij}(s)I_{j}(t,s)
\end{equation}

\begin{equation}
\frac{\partial}{\partial t}I_{i}(t,s)+\frac{\partial}{\partial s}I_{i}(t,s)=0
\end{equation}

\begin{equation}
I_{i}(t,-1)=R_{0,i}S_{i}\sum_{j=1}^{M}\int_{-1}^{1}ds\beta_{ij}(s)I_{j}(t,s)
\end{equation}

\begin{equation}
\frac{dR_{i}}{dt}=I_{i}(t,1)
\end{equation}

As before, there are many valid discretizations of the PDE model but as before we will focus on the Galerkin
approximation and the method of lines.

\subsection{Discretization: Method of Lines}

Once again, we discretize $s$ in uniform intervals of width $h$ and represent advection/integration
in $s$ via upwinding and left Riemann sums, respectively. We evaluate the solution at discrete points
in s, evaluating at $s_{1},s_{2},s_{3}\ldots,s_{N}$ where $s_{n}=-1+(n-1)h$ and $h=2/(N-1)$. In this
section, we denote $I_{i,n}$ as the number density of infected persons at each $s_{n}$ and $\beta_{ij,n}=h\beta_{ij}(s_{n})$.
Applying this discretization scheme to the PDE equations, we obtain evolution equations for $S_{i},I_{i,n},$
and $R_{i}$:

\begin{equation}
\frac{dS_{i}}{dt}=-R_{0,i}S_{i}\sum_{n=2}^{N}\beta_{ij,n}I_{i,n}
\end{equation}

\begin{equation}
I_{i,0}=R_{0,i}S_{i}\sum_{n=2}^{N}\beta_{ij,n}I_{i,n}
\end{equation}

\begin{equation}
\frac{dI_{i,n}}{dt}=\frac{1}{h}(I_{i,n-1}-I_{i,n})
\end{equation}

\begin{equation}
\frac{dR_{i}}{dt}=I_{i,N}
\end{equation}

This maps exactly to the standard age-structured multi-stage SIkR model. Therefore, we confirm that in
the limit of $h\rightarrow0$, the age-structured SIkR model is equivalent to an age-structured time
since infection model. In practice, however, the convergence is so slow that truncation errors will have
a large influence on the predictions. As before, the truncation error can be represented as numerical
diffusion in the $s-$direction.

As before, one can also employ the more efficient predictor/corrector discretization using the method
of lines.

\subsection{Discretization: Galerkin Approximation}

We can represent each $I_{i}(t,s)$ on $s\in[-1,1]$ via a truncated Legendre polynomial expansion with
$m+1$ polynomials:

\begin{equation}
I_{i}(t,s)=\sum_{n=0}^{m}I_{i,n}(t)P_{n}(s)
\end{equation}
Inserting this into the PDE equations, we obtain:

\begin{equation}
\frac{dS_{i}}{dt}=-R_{0,i}S_{i}\sum_{j=1}^{M}\sum_{n=0}^{m}a_{ij,n}I_{j,n}
\end{equation}

\begin{equation}
a_{ij,n}=\int_{-1}^{1}ds\beta_{ij}(s)P_{n}(s)
\end{equation}

\begin{equation}
\frac{d}{dt}I_{i,n}+\sum_{k=0}^{m}b_{nk,i}I_{i,n}=0
\end{equation}

The coefficients $b_{nk,i}$ evaluate to $b_{nk}=2$n+1 if $n+k$ is odd and $k>n$, otherwise they evaluate
to zero. The highest order Legendre Polynomial coefficient $I_{i,m}$is constrained to satisfy:

\begin{equation}
\sum_{n=0}^{m}I_{i,n}(-1)^{n}=R_{0,i}S_{i}\sum_{j=1}^{M}\sum_{n=0}^{m}a_{ij,n}I_{j,n}
\end{equation}

\subsection{Linear Stability Analysis}

As a generalization of the linear stability analysis for the single compartment case, we suppose an initial
condition in which nearly the whole population is susceptible and there is a small population of infected
growing as $I_{i}(t,s)=I_{0,i}\exp(\lambda(t-s))$. The growth rate $\lambda$ and relative number of
infected in each age compartment $I_{0,i}$ can be obtained by considering the following matrix:

\begin{equation}
A_{ij}=R_{0,i}S_{i}\sum_{j=1}^{M}\int_{-1}^{1}ds\beta_{ij}(s)e^{-\lambda(s+1)}
\end{equation}

The growth rate $\lambda$ is the value of lambda for which the largest eigenvalue of $A_{ij}$ is equal
to one, and the relative number of infected in each age compartment $I_{0,i}$ is given by the corresponding
eigenvector.
\end{document}